\documentclass[epj]{svjour}
\usepackage{graphics}
\usepackage{bm}
\usepackage{bbold}
\usepackage{amssymb}
\usepackage{hyperref}

\renewcommand{\vec}[1]{{\mathbf #1}}

\usepackage{color}
\newcommand{\red}[1]{#1}

\begin{document}
\title{Finite-size scaling of the density of states inside band gaps of ideal and disordered photonic crystals}
\author{Sergey E. Skipetrov
}                     
%
%
\institute{Univ. Grenoble Alpes, CNRS, LPMMC, 38000 Grenoble, France}
\date{Received: date / Revised version: date}
%
\abstract{
We study the density of states (DOS) in band gaps of ideal and disordered three-dimensional photonic crystals of finite size. The ideal crystal is a diamond lattice of resonant point scatterers (atoms) whereas the disordered one is obtained from it by displacing the scatterers by random distances in random directions. We find that DOS inside a band gap of the ideal crystal decreases as the inverse of the crystal size. Disorder narrows the band gap and DOS exhibits enhanced fluctuations near the new band edges. However, the average DOS still exhibits the same scaling with the crystal size within the remaining band gap. A phenomenological  explanation of this scaling suggests that it should hold for one- and two-dimensional photonic crystals as well.
\PACS{
      {42.70.Qs}{Photonic band gap materials}   \and
      {78.67.Pt}{Optical properties of photonic structures} \and
      {42.25.Dd}{Wave propagation in random media}
     } 
} 

\authorrunning{S.E. Skipetrov}
\titlerunning{Finite-size scaling of DOS inside band gaps}

\maketitle
\section{Introduction}
\label{intro}

Photonic crystals are promising for many practical applications and constitute an interesting physical object to study \cite{joan08}. Theoretical studies of their optical properties are largely facilitated by their ideal periodic structure and a common assumption of infinite crystal size. However, the crystals used in current experiments and to be used in future potential applications are necessarily imperfect \cite{koen05,ton08} and of finite size. These complications are not minor and have important consequences that can spoil the important physical properties that make photonic crystals interesting. In particular, both disorder and finite size introduce states (modes) inside a band gap of a hypothetical ideal infinite crystal---a phenomenon that is clearly undesirable for those applications that rely on the vanishing density of optical states (DOS) inside a band gap.

Scaling of DOS inside a band gap with the size $L$ of an ideal photonic crystal has been theoretically studied by Bin Hasan \textit{et al.} \cite{bin18}. These authors have predicted that inside a band gap, DOS is inversely proportional to $L$. However, their analysis was based on a phenomenological assumption that optical modes of the crystal acquire finite lifetimes proportional to $L$. A different dependence of mode lifetimes on $L$ would lead to a different result. In addition, different modes are not necessarily affected by the finite crystal size in the same way and thus their lifetimes may be different. Finally, real crystals are not only finite in size but are also disordered due to imperfections that inevitably arise during the fabrication process \cite{koen05,ton08}. Therefore, the question concerning the scaling of DOS with crystal size in realistic, disordered photonic crystals cannot be considered as completely closed.

The authors of Ref.\ \cite{bin18} write that ``exact calculations for 2D and 3D photonic crystals are beyond the scope of present-day computational power''. In this paper we use the model of dipolar resonant point scatterers (atoms) arranged in a 3D diamond lattice to challenge this statement and study the scaling of DOS with the size $L$ of photonic crystal from first principles. In contrast to the calculation of Ref.\ \cite{bin18}, our model is fully microscopic. It can be realized in an experiment with cold atoms in an optical lattice and serve as a qualitative model of photonic crystals made of identical small dielectric particles. We show that, indeed, DOS scales as $1/L$ inside a band gap of a perfect crystal. In addition, we demonstrate that introducing disorder in scatterer positions reduces the gap but does not affect this scaling as far as disorder does not close the gap. Deviations from $1/L$ scaling of DOS appear near band edges where DOS exhibits large fluctuations from one realization of disorder to another.

\section{The model}
\label{sec:model}

\begin{figure*}[t]
\resizebox{1\textwidth}{!}{%
\includegraphics{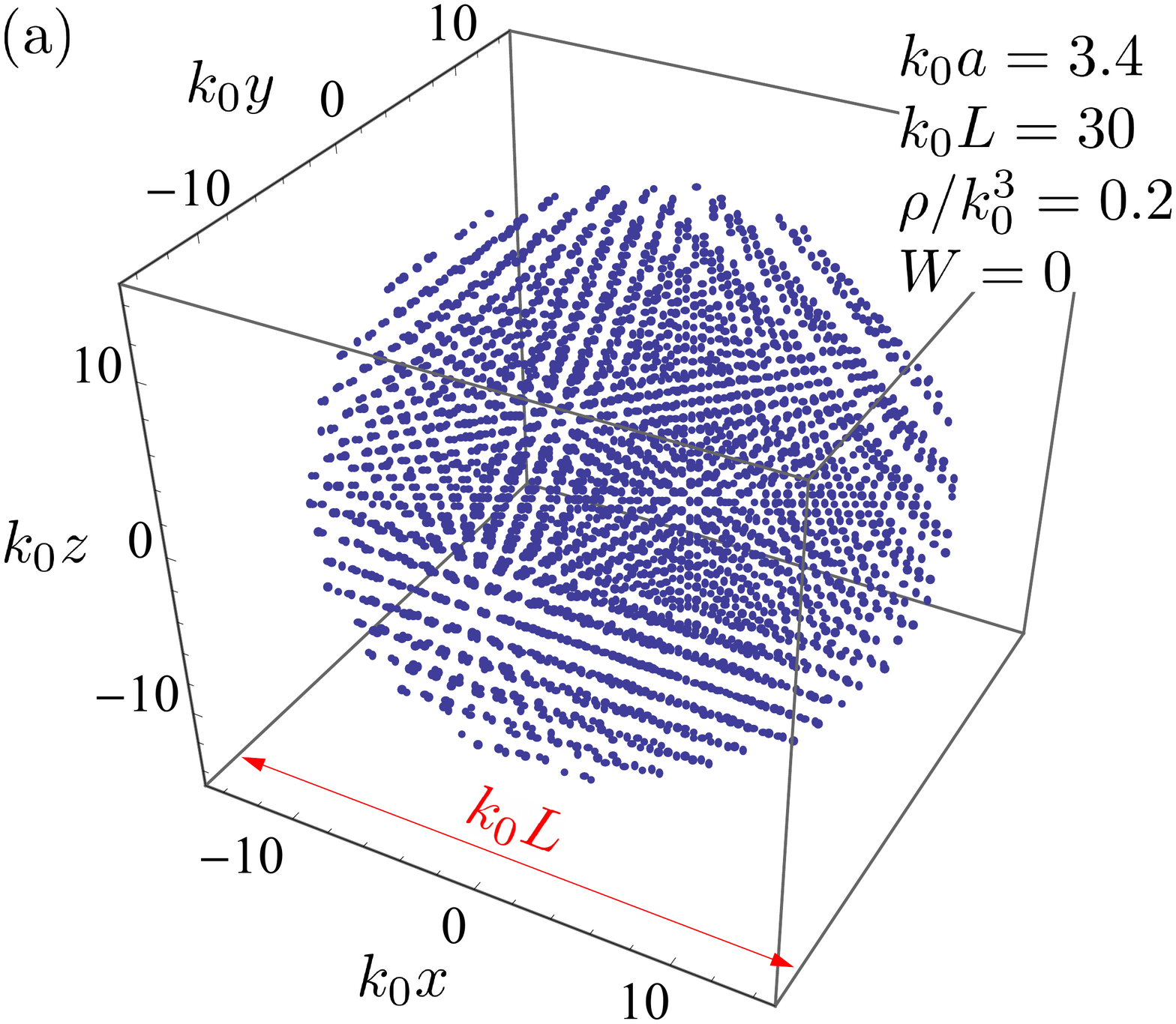}\\
\includegraphics{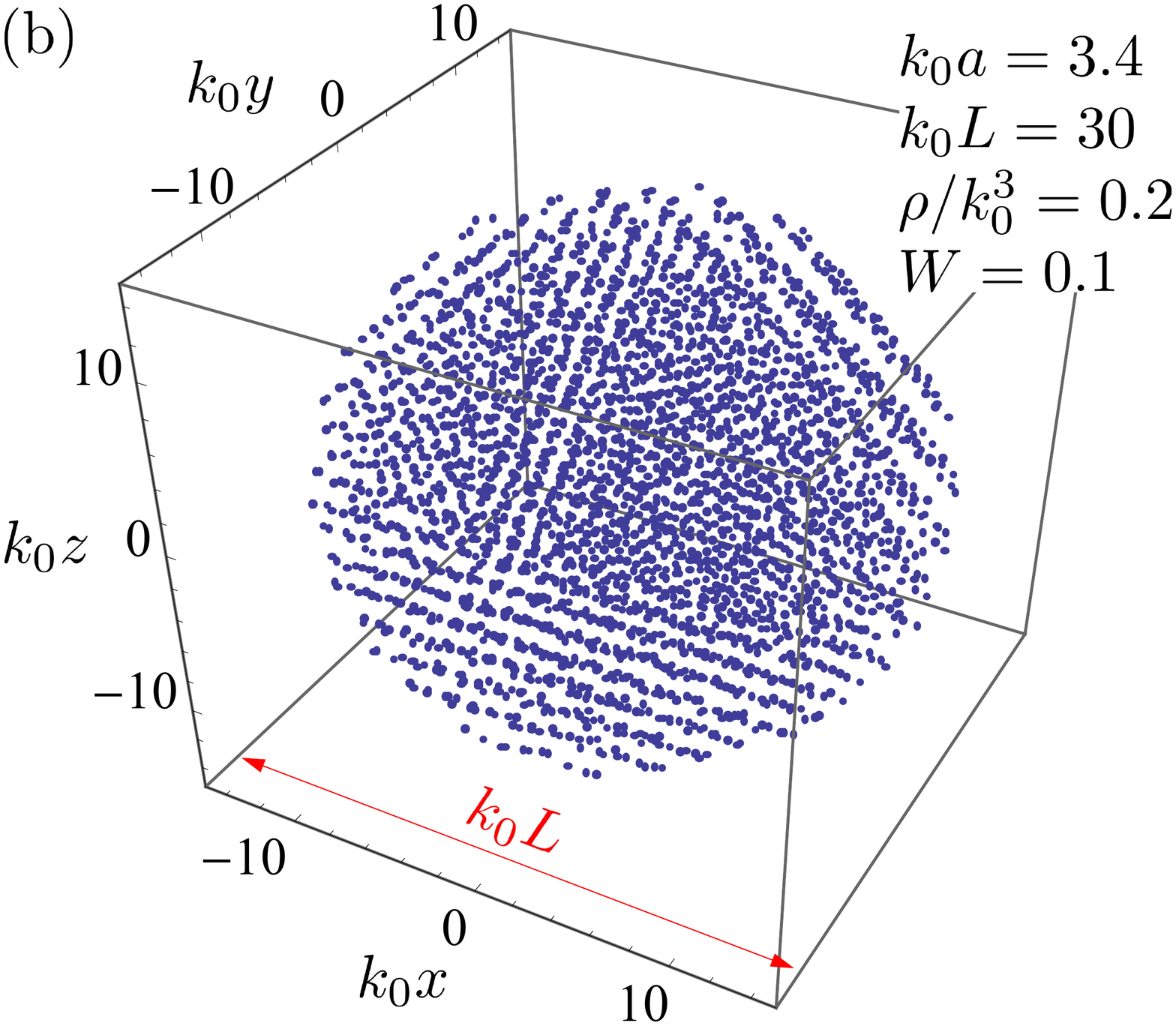}\\
}
\caption{Visualization of perfect (a) and disordered (b) three-dimensional diamond lattices studied in this work. Blue points indicate positions of resonant point scatterers (atoms) with a resonance frequency $\omega_0$ and a resonance width $\Gamma_0$. The perfect lattice (a) is obtained by superposition of two identical fcc lattices with a lattice constant $a$. The basis vectors of the first lattice are $\vec{e}_1 = (0, a/2, a/2)$,  $\vec{e}_2 = (a/2, 0, a/2)$ and  $\vec{e}_3 = (a/2, a/2, 0)$. The second lattice is obtained by translating the first one by $\vec{e} = (a/4, a/4, a/4)$. $k_0 a = 3.4$ for this figure with $k_0 = \omega_0/c$ and $c$ the speed of light. The lattice spacing $a$ and the diameter $L$ of the system fix the number $N$ and the average number density $\rho$ of scatterers ($N = 2869$ and $\rho/k_0^3 = 0.2$ for this figure). The disordered lattice (b) is obtained from (a) by displacing scatterers in random directions by random distances $\Delta r_j$ ($j = 1, \ldots, N$) that are uniformly distributed within an interval $[0, \Delta r_{\mathrm{max}}]$, where $\Delta r_{\mathrm{max}} = W \times a$ and $W$ measures the strength of disorder. $W = 0.1$ for the panel (b) of this figure.}
\label{fig:crystal}
\end{figure*}

We study a periodic arrangement of $N \gg 1$ identical dipolar resonant point scatterers (atoms) having a resonance frequency $\omega_0$ and a resonance width $\Gamma_0$. Previous studies including pioneering works by Foldy \cite{foldy45} and Lax \cite{lax51} as well as later developments \cite{agarwal70,lehmberg70,rusek96,sokolov11} have shown that the multiple scattering of light in a system of $N$ point scatterers can be studied using a Green's matrix ${\hat G}$ that plays a role of an effective Hamiltonian for the atomic subsystem of the full system ``atoms $+$ light''. The $3N \times 3N$ matrix ${\hat G}$ is composed of $N \times N$ blocks of size $3 \times 3$, each block describing propagation of light between a pair of atoms. The explicit expression for the elements of a $3 \times 3$ block ${\hat G}_{jn}$ of the matrix ${\hat G}$ is
\begin{eqnarray}
G_{jn}^{\mu \nu} &=& i\delta_{jn} \delta_{\mu \nu}
+ (1 - \delta_{jn}) \frac{3}{2}
\frac{e^{i k_0 r_{jn}}}{k_0 r_{jn}}
\nonumber \\
&\times& \left[ P(i k_0 r_{jn}) \delta_{\mu \nu}
+ Q(i k_0 r_{jn})
\frac{r_{jn}^{\mu} r_{jn}^{\nu}}{(r_{jn})^2} \right],
\end{eqnarray}
where $P(x) = 1-1/x+1/x^2$, $Q(x) = -1+3/x-3/x^2$, $\vec{r}_{jn}$ is a vector connecting atoms $j$ and $n$ ($j$, $n = 1, \ldots, N$), $k_0 = \omega_0/c$, $c$ is the speed of light in the free space, and the indices $\mu$, $\nu = x, y, z$ denote the projections of $\vec{r}_{jn}$ on the axes of the coordinate system: $r_{jn}^{x} = x_{jn}$, $r_{jn}^{y} = y_{jn}$, $r_{jn}^{z} = z_{jn}$.

The matrix ${\hat G}$ allows for calculating the vector $\vec{E}(\omega)$ containing three projections of $N$ electric field vectors at the positions of atoms due to an incident monochromatic wave $\vec{E}_0(\omega)$ via an equation \cite{foldy45,lax51}:
\begin{eqnarray}
\vec{E}(\omega) = \vec{E}_0(\omega) + \alpha(\omega) \left[ {\hat G}(\omega) - i \mathbb{1} \right] \vec{E}(\omega),
\label{foldylax}
\end{eqnarray}
where $\vec{E}(\omega) =[E^x(\omega, \vec{r}_1)$, $E^y(\omega, \vec{r}_1)$, $E^z(\omega, \vec{r}_1)$, $\ldots$, $E^x(\omega,$ $\vec{r}_N)$, $E^y(\omega, \vec{r}_N)$, $E^z(\omega, \vec{r}_N)]^{T}$, $\vec{E}_0(\omega) = [E_0^x(\omega, \vec{r}_1)$, $E_0^y(\omega,$ $\vec{r}_1)$, $E_0^z(\omega, \vec{r}_1)$, $\ldots$, $E_0^x(\omega, \vec{r}_N)$, $E_0^y(\omega, \vec{r}_N)$, $E_0^z(\omega, \vec{r}_N)]^{T}$ and $\alpha(\omega) = -(\Gamma_0/2)/(\omega - \omega_0 + i \Gamma_0/2)$ is the (dimensionless) scatterer polarizability. The solution of Eq.\ (\ref{foldylax}) reads
\begin{eqnarray}
\vec{E}(\omega) = \left( \mathbb{1} - \alpha(\omega) \left[ {\hat G}(\omega) - i \mathbb{1} \right] \right)^{-1}
\vec{E}_0(\omega).
\label{sol}
\end{eqnarray}

It follows from Eq.\ (\ref{foldylax}) that any field $\vec{E}(\omega)$ can be expanded in $3 N$ right eigenvectors
\red{
$\bm{\psi}_m = (\psi_m^{1}, \ldots, \psi_m^{3N})^T$
}
of the matrix ${\hat G}$ obeying
\begin{eqnarray}
{\hat G} \bm{\psi}_m = \Lambda_m \bm{\psi}_m,\;\;\;\; m = 1,\ldots,3N,
\label{eigen}
\end{eqnarray}
where the eigenvalues $\Lambda_m$ are complex due to the nonhermiticity of the matrix ${\hat G}$. Following the usual terminology of the field of open quantum system, we will call $\bm{\psi}_m$ quasimodes. Then $\omega_m = \omega_0 - (\Gamma_0/2) \mathrm{Re} \Lambda_m$ and $\Gamma_m = \Gamma_0 \mathrm{Im} \Lambda_m$ represent the oscillation frequencies and decay rates of quasimodes, respectively.
\red{
We note for completeness that the matrix ${\hat G}$ also has a set of left eigenvectors $\bm{\phi}_m$ obeying
\begin{eqnarray}
\bm{\phi}_m {\hat G} = \bm{\phi}_m \Lambda_m,\;\;\;\; m = 1,\ldots,3N.
\label{eigenleft}
\end{eqnarray}
Right and left eigenvectors form a complete biorthogonal basis \cite{morse53}:
\begin{eqnarray}
\sum\limits_{j=1}^{3N} \phi_m^{j} \psi_n^{j} = \delta_{mn}.
\label{biorth}
\end{eqnarray}
}

\begin{figure*}[t]
\resizebox{1\textwidth}{!}{%
\includegraphics{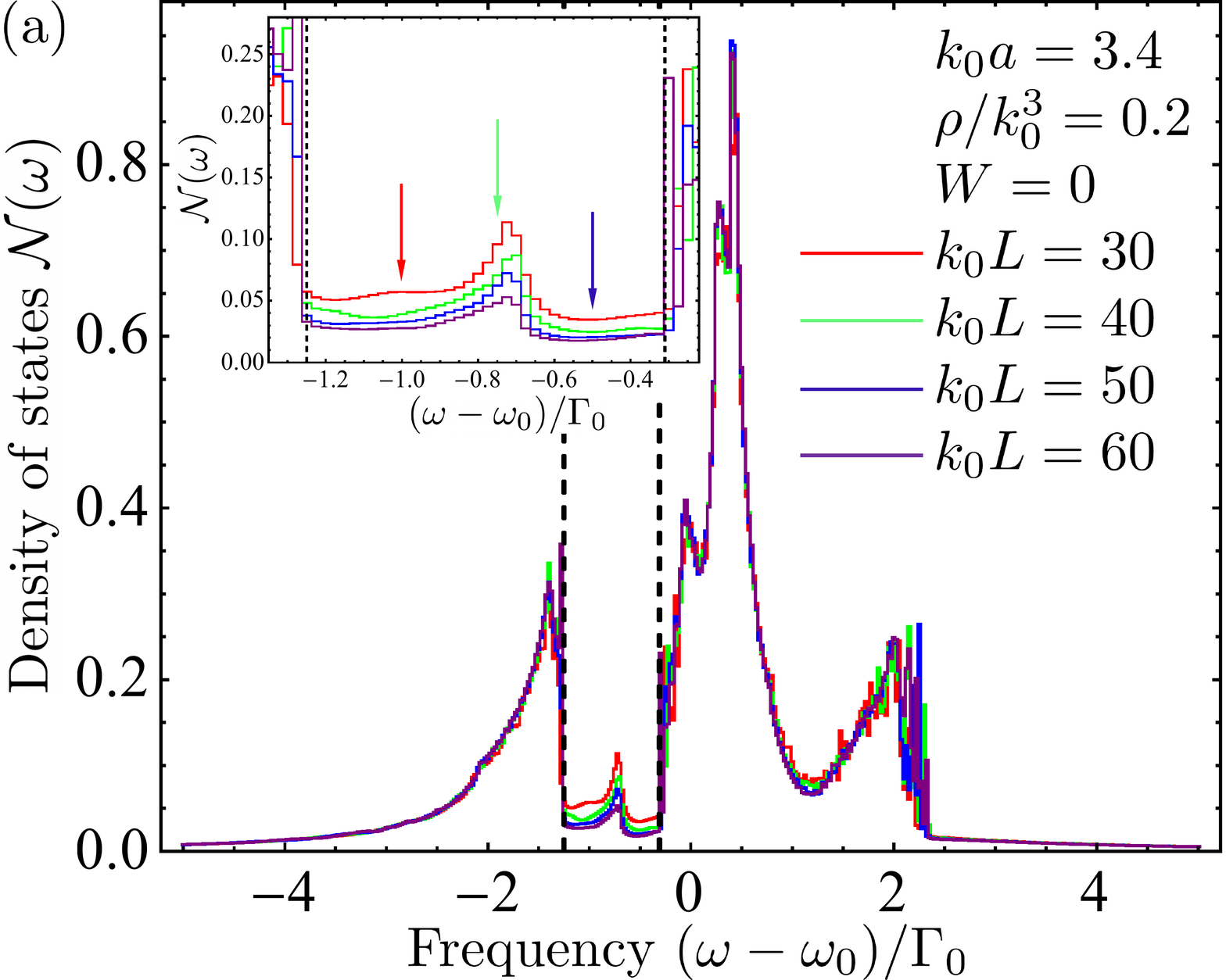}\\
\includegraphics{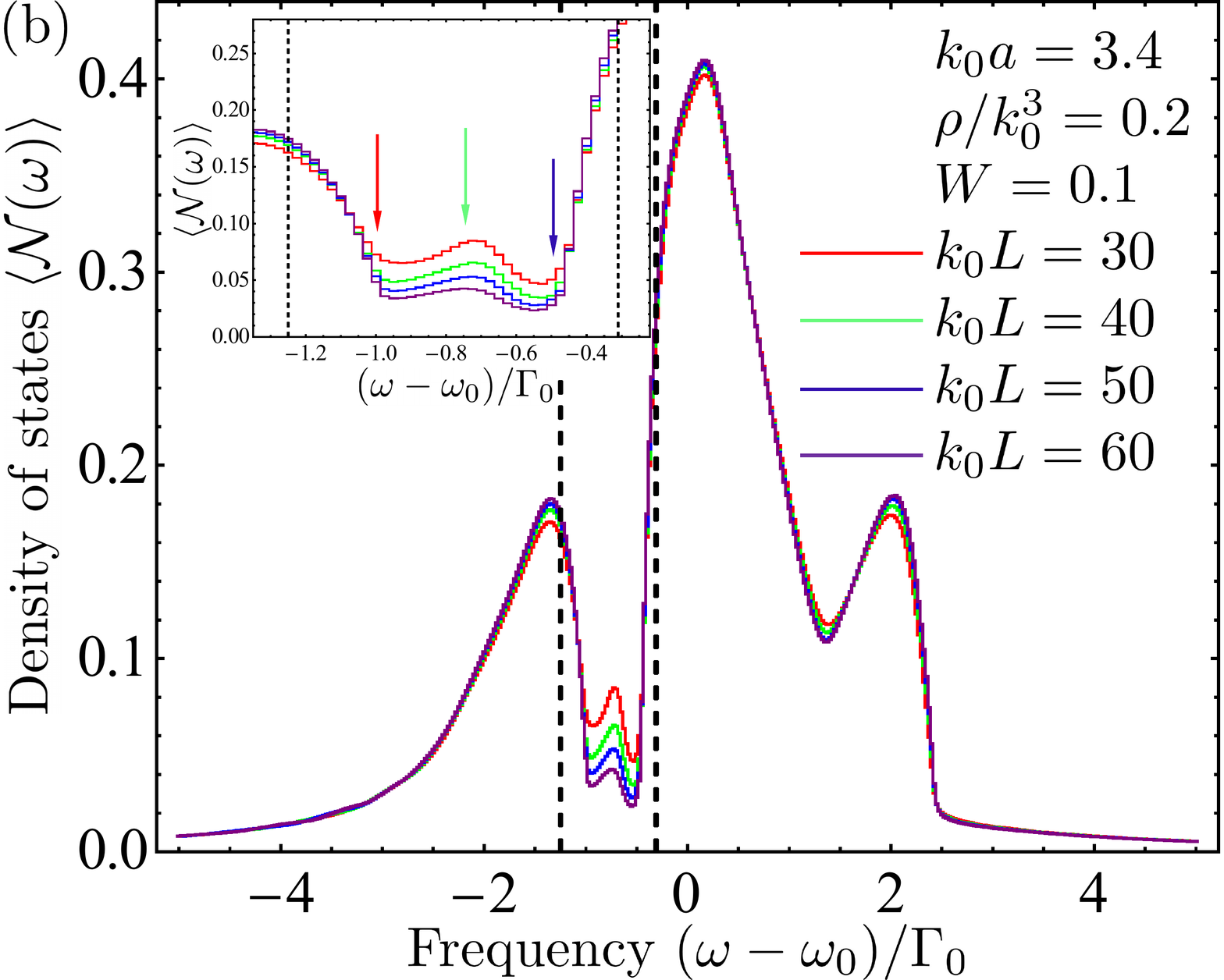}\\
}
\caption{
DOS (a) and average DOS (b) of the perfect and disordered crystals shown in Figs.\ \ref{fig:crystal}(a) and (b), respectively, for an increasing crystal size $L$. In (b), DOS is averaged over
\red{1900, 800, 436 and 109}
independent random scatterer configurations for $k_0 L = 30$, 40, 50 and 60, respectively.
Dashed lines show band edges as determined in Ref.\ \cite{antezza09} for the infinite crystal.
Insets zoom on the spectral range corresponding to the band gap.
Arrows in the inset indicate frequencies for which scaling with $L$ is shown in Fig.\ \ref{fig:scaling}.
}
\label{fig:dos}
\end{figure*}

In the past, the model presented above has been used to study light scattering in fully disordered systems by assuming that atomic positions $\vec{r}_j$ are random and independent \cite{rusek96,fofanov13,skip14}, in photonic crystals \cite{antezza13}, and more recently in aperiodic systems \cite{sgri19,sgri19ol}. We start with the same diamond lattice as in Ref.\ \cite{antezza13} but introduce disorder by randomly displacing all atoms instead of taking some of them out. The diamond lattice is obtained by superimposing two identical face-centered cubic (fcc) lattices with a lattice constant $a$. The basis vectors of the first lattice are $\vec{e}_1 = (0, a/2, a/2)$,  $\vec{e}_2 = (a/2, 0, a/2)$ and  $\vec{e}_3 = (a/2, a/2, 0)$. The second lattice is obtained by translating the first one by $\vec{e} = (a/4, a/4, a/4)$. An example of an ideal diamond lattice obtained in this way is shown in Fig.\ \ref{fig:crystal}(a). Disorder is introduced into the ideal lattice by displacing atoms by random distances $\Delta r_j$ in random, independent directions. We choose $\Delta r_j$ uniformly distributed between 0 and $W \times a$, where the dimensionless quantity $W$ measures the strength of disorder. An example of a disordered atomic lattice is shown in Fig.\ \ref{fig:crystal}(b).

The ideal diamond lattice exhibits a full omnidirectional band gap for $k_0 a \lesssim 5.14$ \cite{antezza09}. In this work we choose a fixed value of $k_0 a = 3.4$ corresponding to the average number density of scatterers $\rho = N/V \simeq 0.2$, where $V$ is the volume of the crystal. To avoid any artifacts that may arise from the presence of corners in a finite-size crystal (for example, for a crystal having a shape of a cube) or from a particular symmetry of crystal shape coinciding with the symmetry of the crystal structure (for a diamond-shaped crystal), we assume that the crystalline structure fills a sphere of diameter $L$ (see Fig.\ \ref{fig:crystal}).

\section{Density of states}
\label{sec:dos}

\subsection{Definitions}
\label{sec:definition}

\red{
In any medium, the local density of optical states (LDOS) at a point $\vec{r}$ and at a frequency $\omega$, ${\cal N}(\omega, \vec{r})$, can been shown to be proportional to the imaginary part of the trace of the dyadic Green's function describing the electric field generated by a harmonically oscillating point dipole at a point $\vec{r}$, evaluated at $\vec{r}$ \cite{joulain03,colas01,chicanne03}.
DOS ${\cal N}(\omega)$ in a sample of finite size is given by the spatial integral of ${\cal N}(\omega, \vec{r})$ over the volume of the sample. In order to evaluate DOS in the photonic crystals considered in this work, one has to solve Eq.\ (\ref{foldylax}) for $\vec{E}(\omega)$ with $\vec{E}_0(\omega)$ being the field emitted by a point dipole at $\vec{r}$ and then evaluate the imaginary part of the field at $\vec{r}$ resulting from the interference of fields scattered by all atoms with the emitted field $\vec{E}_0(\omega)$.
Although statistical properties of ${\cal N}(\omega, \vec{r})$ has been studied using Eq.\ (\ref{foldylax}) for a single point inside a disordered medium and a single frequency \cite{caze10}, integrating the result over the volume of the sample and repeating the analysis for different frequencies constitutes a formidable computational task and falls beyond the scope of this work. Instead, we notice that if we assume that an atom $j$ of our ensemble of $N$ atoms is the only one to be excited by the incident field $\vec{E}_0$, its polarization $\vec{P} = \alpha(\omega) \vec{E}(\omega)$ can be expanded in right $\bm{\psi}_m$ and left $\bm{\phi}_m$ eigenvectors of the matrix ${\hat G}$ as
\begin{eqnarray}
P^{\mu\nu}(\omega, \vec{r}_j) \propto -\sum\limits_{m=1}^{3N}
\frac{\phi_m^{3(j-1)+\mu} \psi_m^{3(j-1)+\nu}}{\omega - \omega_m + i \Gamma_m/2},
\label{expansion}
\end{eqnarray}
where $P^{\mu\nu}$ denotes the polarization component $\nu$ due to the incident field along $\mu$ axis and $\mu, \nu = 1,2,3$ for $x$, $y$, $z$ axes, respectively.
LDOS is then found as the imaginary part of the trace of the response to the point-like excitation:
\begin{eqnarray}
{\cal N}(\omega, \vec{r}_j) &\propto& \mathrm{Im} \left[ \mathrm{Tr}\; P^{\mu\nu}(\omega, \vec{r}_j) \right]
\nonumber \\
&\propto&  \mathrm{Im} \sum\limits_{\mu = 1}^3 \sum\limits_{m=1}^{3N}
\frac{\phi_m^{3(j-1)+\mu} \psi_m^{3(j-1)+\mu}}{\omega - \omega_m + i \Gamma_m/2}.
\label{ldos}
\end{eqnarray}
Finally, DOS is obtained by summing over all atoms $j$:
\begin{eqnarray}
{\cal N}(\omega) &=& \frac{1}{N} \sum\limits_{j=1}^N {\cal N}(\omega, \vec{r}_j)
\nonumber \\
&\propto&
\sum\limits_{m=1}^{3N}
\frac{(\Gamma_m/2)}{(\omega - \omega_m)^2 + (\Gamma_m/2)^2},
\label{dos1}
\end{eqnarray}
where we have used the biorthogonality condition (\ref{biorth}).
}

\begin{figure*}[t]
\resizebox{1\textwidth}{!}{%
\includegraphics{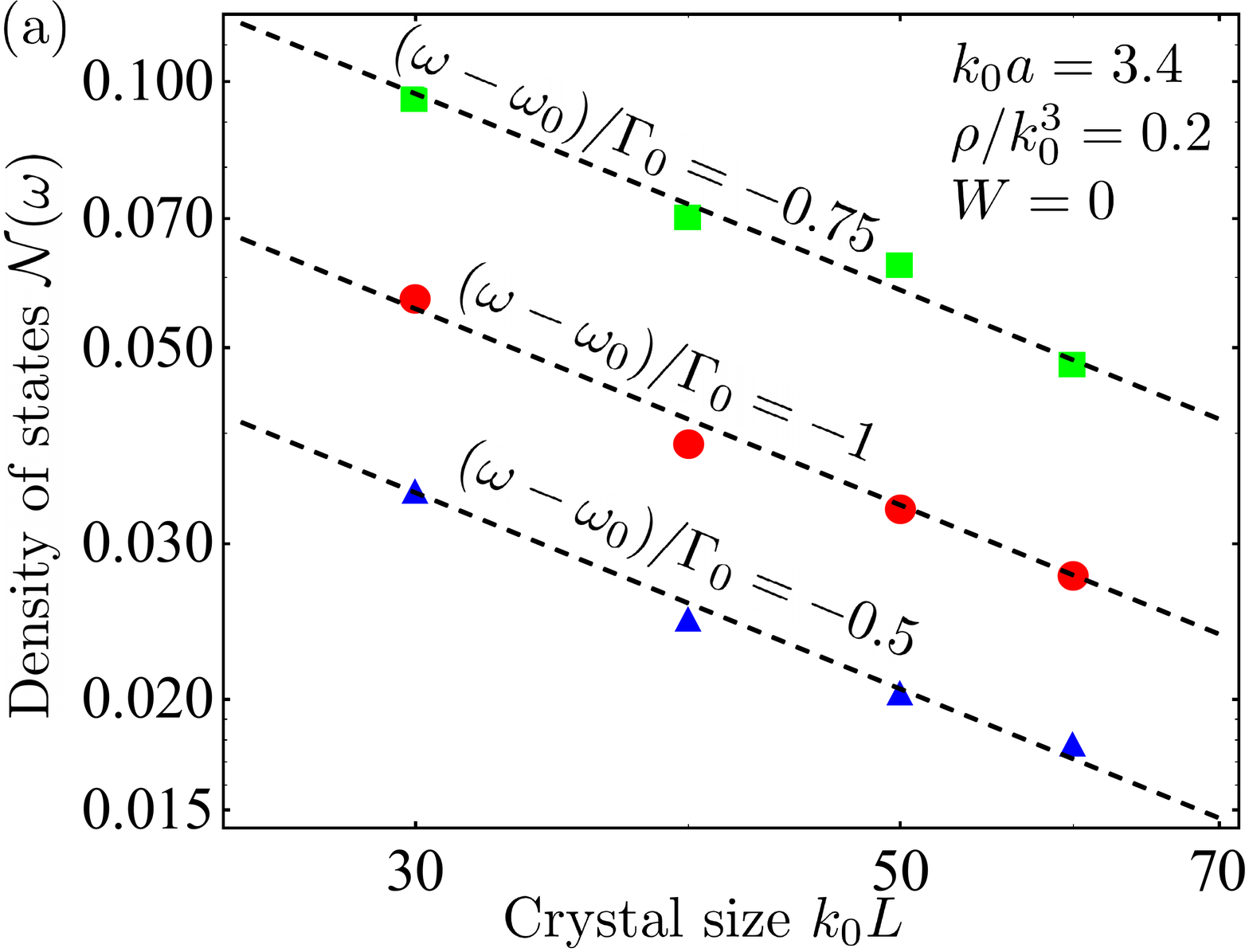}\\
\includegraphics{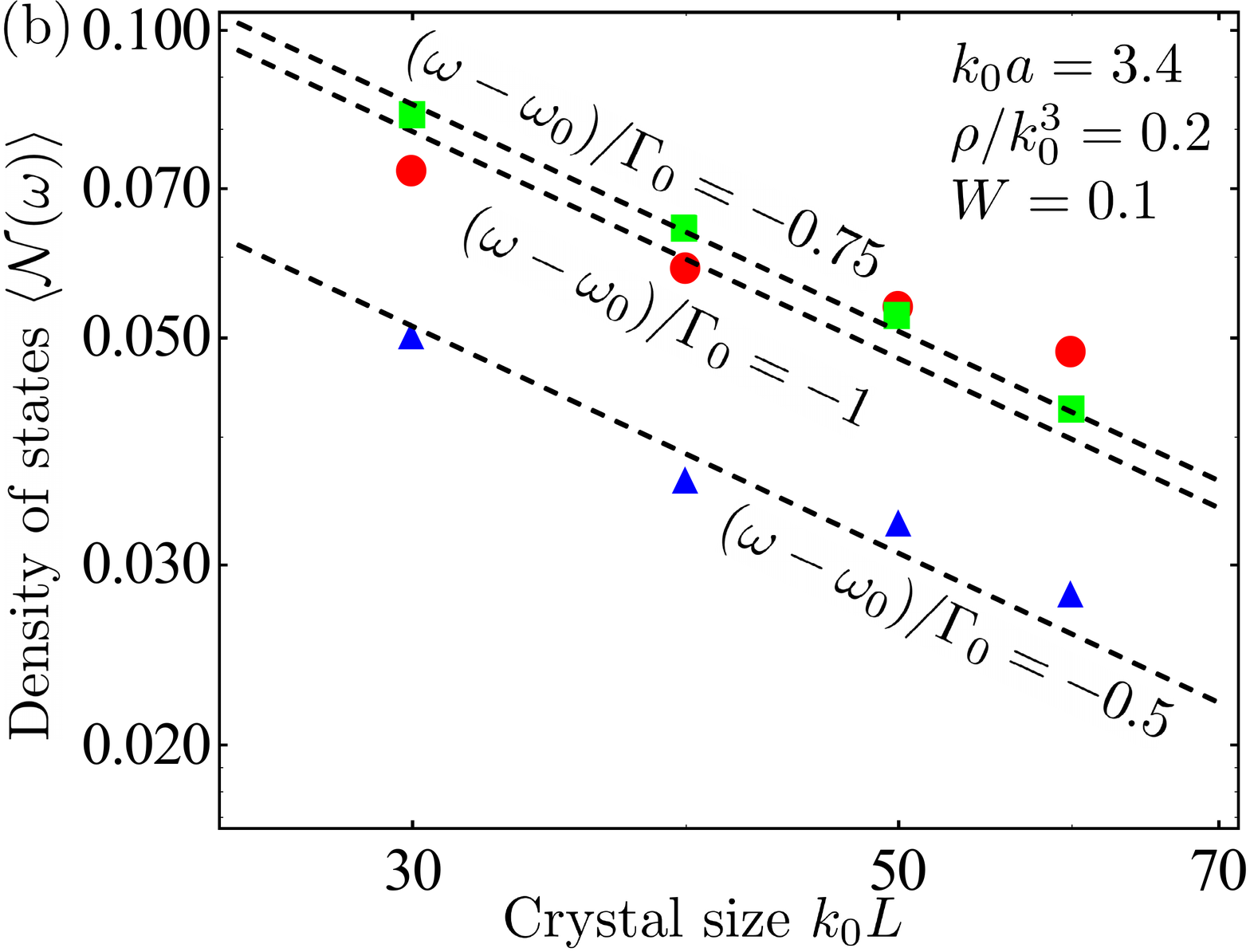}\\
}
\caption{
Scaling of DOS (a) and of average DOS (b) shown in Fig.\ \ref{fig:dos}, with crystal size $L$, for three frequencies indicated in the insets of Figs.\ \ref{fig:dos}(a) and (b) by arrows.
Dashed lines show power-law fits ${\cal N}(\omega)$, $\langle {\cal N}(\omega) \rangle \propto 1/L$.
}
\label{fig:scaling}
\end{figure*}

In order to compare results corresponding to different $N$, we normalize DOS to one:
\begin{eqnarray}
\int\limits_{-\infty}^{\infty} d\omega {\cal N}(\omega) = 1,
\label{dosnorm}
\end{eqnarray}
so that the number of states in an infinitesimal frequency interval $d\omega$ centered at a frequency $\omega$ is $3N {\cal N}(\omega) d\omega$.
\red{
This leads to the final, normalized expression for DOS that we will use the following:
}
\begin{eqnarray}
{\cal N}(\omega) = \frac{1}{3 N \pi}\sum\limits_{m=1}^{3N} \frac{(\Gamma_m/2)}{(\omega - \omega_m)^2 + (\Gamma_m/2)^2}.
\label{dos}
\end{eqnarray}
For a closed or infinite system $\Gamma_m \to 0$ and Eq.\ (\ref{dos}) reduces to the familiar expression
$(1/3N)\sum_{m=1}^{3N} \delta(\omega - \omega_m)$.

\red{
It is important to note that elementary excitations of the system ``atoms + light'' are a kind of polaritons having both atomic and electromagnetic field components. Equation (\ref{dos}) takes into account only the atomic component and thus has two important limitations. First, it does not approach DOS of the free electromagnetic field for small $N$ or low atomic densities. For a single atom ($N = 1$), for example, $\omega_m = \omega_0$ and $\Gamma_m = \Gamma_0$, so that Eq.\ (\ref{dos}) reduces to a simple Lorentzian centered at $\omega = \omega_0$. Its decay with $|\omega - \omega_0|$ simply reflects the fact that the atom-field interaction weakens when $\omega$ moves away from $\omega_0$ and hence the atomic component of the polariton diminishes. Second, Eq.\ (\ref{dos}) completely misses those excitations which has no atomic component at all and vanish exactly at atomic positions $\{ \vec{r}_j \}$. These so-called free-field solutions have been already discussed in Refs.\ \cite{klug06,antezza09} and have been shown to have frequencies $\omega_{\mathrm{free}} > \omega_0$. Therefore, they do not affect DOS inside spectral gaps found for $\omega < \omega_0$ as we will see in Secs.\ \ref{sec:ideal} and \ref{sec:disordered}. Clearly, the two limitations of Eq.\ (\ref{dos}) discussed above preclude its use for quantitative calculations of DOS but do not affect its ability to predict the {\it scaling} of DOS with the sample size inside spectral gaps.
}

\subsection{Ideal diamond lattice}
\label{sec:ideal}

Figure\ \ref{fig:dos}(a) shows DOS of the ideal diamond lattice calculated by numerically diagonalizing matrices ${\hat G}$ of increasing sizes $L$ and then using Eq.\ (\ref{dos}). ${\cal N}(\omega)$ exhibits quite an irregular structure with a number of sharp peaks and regions of strong oscillations. It clearly shows a significant drop in the range $(\omega-\omega_0)/\Gamma_0 \in [-1.25, -0.31]$ corresponding to the band gap of the infinite crystal \cite{antezza09} and delimited by vertical dashed lines. Outside the band gap, DOS changes very little with the crystal size $L$ and seems to have roughly converged to its value in the infinite crystal. In contrast, inside the bandgap DOS clearly decreases with $L$ which we illustrate in the inset of Fig.\ \ref{fig:dos}(a) by showing a zoom on the spectral range corresponding to the band gap. DOS is not flat inside the band gap and exhibits, in particular, a peak around the middle of the band gap at $(\omega-\omega_0)/\Gamma_0 \simeq -0.75$. We have chosen three frequencies $(\omega-\omega_0)/\Gamma_0 = -1$, $-0.75$ and $-0.5$ [indicated by vertical arrows in Fig.\ \ref{fig:dos}(a)] inside the gap and show the dependence of DOS at these frequencies on the crystal size in Fig.\ \ref{fig:scaling}(a). The numerical results are satisfactory described by power-law fits ${\cal N}(\omega) \propto 1/L$ for all three frequencies, including the frequency $(\omega-\omega_0)/\Gamma_0 \simeq -0.75$ corresponding to the peak in DOS. This confirms the prediction of Ref.\ \cite{bin18} on a microscopic basis.

\begin{figure}[t]
\resizebox{1\columnwidth}{!}{%
\includegraphics{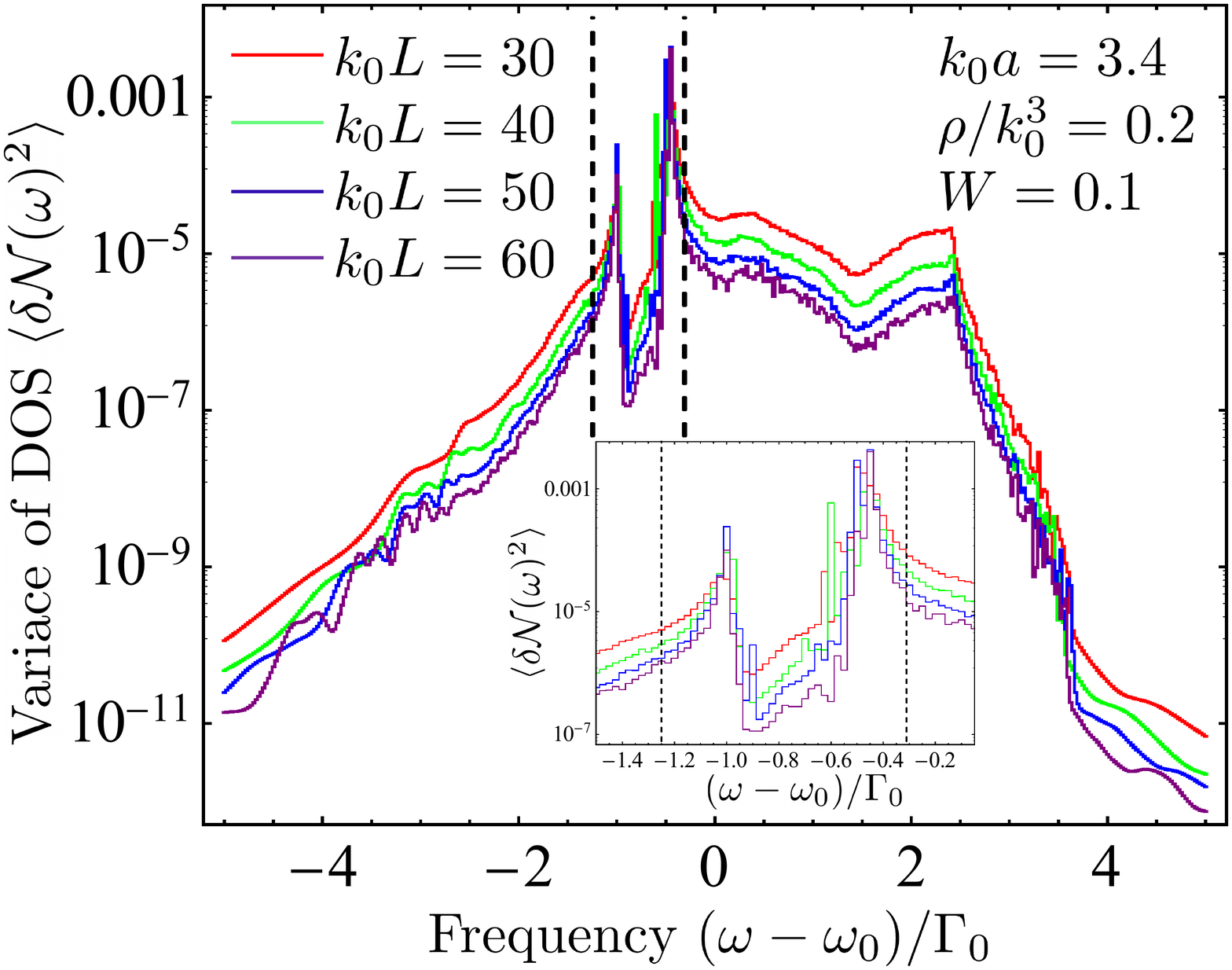}
}
\caption{
Variance of DOS ${\cal N}(\omega)$ in disordered photonic crystals of different sizes $L$. Dashed lines indicate the edges of the band gap in the ideal, infinite crystal without disorder. The inset zooms on the spectral range corresponding to the band gap.}
\label{fig:var}
\end{figure}

\subsection{Disordered diamond lattice}
\label{sec:disordered}

We now turn to the disordered diamond lattice in which atoms are randomly displaced from their positions in an ideal diamond lattice. Because DOS ${\cal N}(\omega)$ of a disordered lattice depends on the exact atomic configuration (i.e. on the realization of ``disorder''), we analyze its average  $\langle {\cal N}(\omega) \rangle$ over an ensemble of many different, statistically independent configurations. This quantity is shown in Fig.\ \ref{fig:dos}(b). We see that averaging over disorder smooths out the rapid variations and oscillations of ${\cal N}(\omega)$ [compare Figs.\ \ref{fig:dos}(a) and (b)] while preserving its overall structure and, in particular, the band gap. Its width, however, is reduced by disorder and its edges are washed out, as might have been expected. The peak of DOS in the middle of the gap is only slightly suppressed by disorder.

Figure\ \ref{fig:scaling}(b) shows the scaling of the average DOS with crystal size $L$ for the same three frequencies as in Fig.\ \ref{fig:scaling}(a). In the middle of the gap at $(\omega-\omega_0)/\Gamma_0 \simeq -0.75$, the average DOS shows the same scaling $\langle {\cal N}(\omega) \rangle \propto 1/L$ as DOS in the ideal crystal without disorder [green squares in Fig.\ \ref{fig:scaling}(b)]. The two other frequencies, however, now turn out to be close to the band edges pushed towards each over by disorder. As a result, the decrease of $\langle {\cal N}(\omega) \rangle$ with $L$ slows down and the simple law $\langle {\cal N}(\omega) \rangle \propto 1/L$ describes it less accurately. The accuracy of our results do not allow concluding with certainty whether the two frequencies $(\omega-\omega_0)/\Gamma_0 = -1$ and $-0.5$ are still inside the band gap of the disordered crystal (in which case, the average DOS $\langle {\cal N}(\omega) \rangle$ should decrease with $L$ and vanish for $L \to \infty$) or not (corresponding to a finite $\langle {\cal N}(\omega) \rangle$ in the limit of $L \to \infty$).

Some information about the positions of band edges in the disordered crystal can be obtained by analyzing fluctuations of DOS from one atomic configuration to another. The magnitude of these fluctuations is measured by the variance of DOS $\langle \delta{\cal N}(\omega)^2 \rangle$, where $\delta{\cal N}(\omega) = {\cal N}(\omega) - \langle {\cal N}(\omega) \rangle$. This quantity is shown in Fig.\ \ref{fig:var}. We clearly observe that $\langle \delta{\cal N}(\omega)^2 \rangle$ decreases with the size $L$ of the disordered crystal for almost all frequencies except for the two frequencies inside the band gap of the ideal crystal: $(\omega-\omega_0)/\Gamma_0 \simeq -1$ and $-0.45$. At these two frequencies, $\langle \delta{\cal N}(\omega)^2 \rangle$ exhibits maxima and seems to remain roughly independent from $L$ or even to grow with it (at least, within the accuracy of our calculations). We believe that this behavior signals band edges of the infinite disordered crystal. Indeed, at a frequency corresponding to such a band edge, we expect large fluctuations of DOS for a finite crystal because a slight shift of the band edge for a given atomic configuration with respect to its average position leads to a large change of DOS from 0 inside the band gap (or, more precisely, a small value $\propto 1/L$) to a finite value outside it. We thus see that the two of the three frequencies analyzed in Fig.\ \ref{fig:scaling}, $(\omega-\omega_0)/\Gamma_0 \simeq -1$ and $-0.5$, turn out to be close to band edges of the disordered crystal. This explains deviations from the scaling $\langle {\cal N}(\omega) \rangle \propto 1/L$ at these two frequencies.

\section{Discussion}
\label{sec:discuss}

An insight into the reasons behind $1/L$ scaling of DOS ${\cal N}(\omega)$ with system size $L$ can be obtained by expressing ${\cal N}(\omega)$ via LDOS ${\cal N}(\omega, \vec{r})$:
\begin{eqnarray}
{\cal N}(\omega) = \frac{1}{V}\int\limits_V d^3\vec{r}\; {\cal N}(\omega, \vec{r}).
\label{dosldos}
\end{eqnarray}
In a photonic crystal, ${\cal N}(\omega, \vec{r})$ is suppressed for frequencies inside the band gap \cite{busch98}.
\red{
LDOS varies widely on the scale of a unit cell but its average over a unit cell $\langle {\cal N}(\omega, \vec{r}) \rangle_{\mathrm{cell}}$ is expected to exhibit a roughly exponential decay}
with the distance to the crystal boundary \cite{asa01,yeg14}:
\begin{eqnarray}
\langle {\cal N}(\omega, \vec{r}) \rangle_{\mathrm{cell}} \simeq \langle {\cal N}(\omega, L/2) \rangle_{\mathrm{cell}} \times \exp\left[-\frac{L/2-r}{\xi(\omega)} \right],
\label{ldosexp}
\end{eqnarray}
where $\langle {\cal N}(\omega, L/2) \rangle_{\mathrm{cell}}$ is the value of the unit-cell-averaged LDOS at the crystal boundary and $\xi(\omega) \red{\gg a}$ is its decay length with the distance to the boundary.
The suppression of LDOS deep inside the photonic crystal can lead to an inhibition of spontaneous emission of an excited atom or molecule placed inside the crystal \cite{lodahl04}.
\red{
Substituting Eq.\ (\ref{ldosexp}) into Eq.\ (\ref{dosldos}) yields}
\begin{eqnarray}
{\cal N}(\omega) = \langle {\cal N}(\omega, L/2) \rangle_{\mathrm{cell}} \times \frac{6 \xi(\omega)}{L}
\label{ldospower}
\end{eqnarray}
for $L \gg \xi(\omega)$. This equation features the $1/L$ scaling observed in Fig.\ \ref{fig:scaling}(a).

\red{
The exponential decay of the unit-cell-averaged LDOS (\ref{ldosexp}) is not necessary to obtain the $1/L$ scaling of DOS in Eq.\ (\ref{ldospower}); any other rapid decay (e.g., a linear decay) would lead to the same result except for the numerical prefactor.
}
The model presented above allows for resolving an apparent contradiction between the exponential scaling of LDOS and the power-law scaling of DOS with the crystal size. It is also easy to verify that the model applies in one- and two-dimensional crystals equally well, yielding ${\cal N}(\omega) \propto 1/L$ independent of space dimensionality.

Because LDOS deep inside the crystal is exponentially small, it cannot be relevant for the power-law scaling of DOS and hence the latter is dominated by the states that are spatially localized near the crystal boundary and only weakly sensitive to the band gap. The number of such states scales as the crystal surface $S \propto L^2$ and their weight in the total DOS scales as $S/V \propto 1/L$, where $V \propto L^3$ is the crystal volume. We thus again recover the same $1/L$ scaling from purely geometrical considerations. Obviously, all the arguments presented above still hold in the presence of weak disorder that does not close the band gap.

\section{Conclusions}
\label{sec:concl}

We used the random Green's matrix approach introduced previously \cite{foldy45,lax51,rusek96,skip14,antezza13} to study the density of states (DOS) in photonic crystals. Although more efficient approaches to DOS calculation exist for ideal crystals of infinite extent \cite{antezza09,busch98}, our method is well tailored to deal with systems of finite size and to explore the impact of imperfections (disorder) of the crystal structure. In a photonic crystal composed of identical resonant scatterers (atoms) arranged in a 3D diamond lattice, a full omnidirectional photonic band gap opens when the lattice constant is sufficiently small. Our calculations confirm the previous prediction about the scaling of the density of states ${\cal N}(\omega)$ inside the band gap with the crystal size $L$: ${\cal N}(\omega) \propto 1/L$. This scaling also holds in the presence of weak disorder in scatterer positions, for frequencies $\omega$ inside the band gap of the disordered system, but becomes only approximate and eventually breaks down for frequencies in the vicinity of band edges. The band edges of a disordered photonic crystal are characterized by strong fluctuations of DOS ${\cal N}(\omega)$ from one realization of disorder to another. The variance of these fluctuations $\langle \delta{\cal N}(\omega)^2 \rangle$ does not decrease with crystal size $L$, in contrast to $\langle \delta{\cal N}(\omega)^2 \rangle$ at other frequencies, where $\langle \delta{\cal N}(\omega)^2 \rangle$ is a decaying function of $L$.

\section*{Acknowledgements}

All the computations presented in this paper were performed using the Froggy platform of the CIMENT infrastructure (\href{https://ciment.ujf-grenoble.fr}{{\tt https://ciment.ujf-grenoble.fr}}), which is supported by the Rhone-Alpes region (grant CPER07\verb!_!13 CIRA) and the Equip@Meso project (reference ANR-10-EQPX-29-01) of the program {\it Investissements d'Avenir} supervised by the {\it Agence Nationale de la Recherche}.

%

\begin{thebibliography}{99}
%
%

\bibitem{joan08}
J.D. Joannopoulos, S.G. Johnson, J.N. Winn, and R.D. Meade,
\textit{Photonic Crystals: Molding the Flow of Light. 2nd ed.} (Princeton Univ. Press, Princeton, 2008)

\bibitem{koen05}
A.F. Koenderink, A. Lagendijk, and W.L. Vos,
Phys. Rev. B \textbf{72}, 153102 (2005)

\bibitem{ton08}
C. Toninelli, E. Vekris, G.A. Ozin, S. John, and D.S. Wiersma,
Phys. Rev. Lett. \textbf{101}, 123901 (2008)

\bibitem{bin18}
S. Bin Hasan, A.P. Mosk, W.L. Vos, and A. Lagendijk,
Phys. Rev. Lett. \textbf{120}, 237402 (2018)

\bibitem{foldy45}
L.L. Foldy,
Phys. Rev. \textbf{67}, 107 (1945)

\bibitem{lax51}
M. Lax,
Rev. Mod. Phys. \textbf{23}, 287 (1951)

\bibitem{agarwal70}
G.S. Agarwal,
Phys. Rev. A \textbf{2}, 2038 (1970)

\bibitem{lehmberg70}
R.H. Lehmberg,
Phys. Rev. A \textbf{2}, 883 (1970)

\bibitem{rusek96}
M. Rusek, A. Orlowski, and J. Mostowski,
Phys. Rev. E \textbf{53}, 4122 (1996)

\bibitem{sokolov11}
I.M. Sokolov, D.V. Kupriyanov, and M.D. Havey,
J. Exp. Theor. Phys. \textbf{112}, 246 (2011)

\bibitem{morse53}
\red{
P.M. Morse and H. Feschbach,
\textit{Methods of Theoretical Physics}
(McGraw-Hill, New York, 1953)
}

\bibitem{fofanov13}
Ya.A. Fofanov, A.S. Kuraptsev, I.M. Sokolov, and M.D. Havey,
Phys. Rev. A \textbf{87}, 063839 (2013)

\bibitem{skip14}
S.E. Skipetrov and I.M. Sokolov,
Phys. Rev. Lett. \textbf{112}, 023905 (2014)

\bibitem{antezza13}
M. Antezza and Y. Castin,
Phys. Rev. A \textbf{88}, 033844 (2013)

\bibitem{sgri19}
F. Sgrignuoli, R. Wang, F. A. Pinheiro, and L. Dal Negro,
Phys. Rev. B \textbf{99}, 104202 (2019)

\bibitem{sgri19ol}
F. Sgrignuoli, M. R\"{o}ntgen, C.V. Morfonios, P. Schmelcher, and L. Dal Negro,
Opt. Lett. \textbf{44}, 375 (2019)

\bibitem{antezza09}
M. Antezza and Y. Castin,
Phys. Rev. A \textbf{80}, 013816 (2009)

\bibitem{joulain03}
\red{
K. Joulain, R. Carminati, J.-P. Mulet, and J.-J. Greffet,
Phys. Rev. B \textbf{68}, 245405 (2003).
}

\bibitem{colas01}
\red{
G. Colas des Francs, C. Girard, J.-C. Weeber, C. Chicanne, T. David, A. Dereux, and D. Peyrade,
Phys. Rev. Lett. \textbf{86}, 4950 (2001)
}

\bibitem{chicanne03}
\red{
C. Chicanne, T. David, R. Quidant, J.C. Weeber, Y. Lacroute, E. Bourillot, A. Dereux, G. Colas des Francs, and C. Girard,
Phys. Rev. Lett. \textbf{88}, 097402 (2002)
}

\bibitem{caze10}
\red{
A. Caz\'{e}, R. Pierrat, and R. Carminati,
Phys. Rev. A \textbf{82}, 043823 (2010)
}

\bibitem{klug06}
\red{
J.A. Klugkist, M. Mostovoy, and J. Knoester,
Phys. Rev. Lett. \textbf{96}, 163903 (2006).
}

\bibitem{busch98}
K. Busch and S. John,
Phys. Rev. E \textbf{58}, 3896 (1998)

\bibitem{asa01}
A.A. Asatryan, K. Busch, R.C. McPhedran, L.C. Botten, C. Martijn de Sterke, and N.A. Nicorovici,
Phys. Rev. E \textbf{63}, 046612 (2001)

\bibitem{yeg14}
E. Yeganegi, A. Lagendijk, A.P. Mosk, and W.L. Vos,
Phys. Rev. B \textbf{89}, 045123 (2014)

\bibitem{lodahl04}
P. Lodahl, A.F. van Driel, I.S. Nikolaev, A. Irman, K. Overgaag, D. Vanmaekelbergh, and W.L. Vos,
Nature \textbf{430}, 654 
(2004)


\end{thebibliography}
%

\end{document}